\documentclass[conference]{IEEEtran}
\usepackage{graphicx,epsfig,epstopdf,amsmath,amssymb,enumerate} 
\usepackage{cite}  
\usepackage{multicol}

\begin{document}
\title{Frequency and Quadrature Amplitude Modulation for 5G Networks}
\author{Shangbin~Wu, Yue Wang, Mohammed Al-Imari and Maziar Nekovee \\
\small{Samsung R\&D Institute UK, Communications House, Staines-upon-Thames, TW18 4QE, United Kingdom.} \\
Email: \{shangbin.wu, yue2.wang, m.al-imari, m.nekovee\}@samsung.com

}
\maketitle
\begin{abstract}
The emerging fifth generation (5G) wireless access network, aiming at providing ubiquitous and high data rate connectivity, is envisaged to deploy large number of base stations with higher density and smaller sizes, where inter-cell interference (ICI) becomes a critical problem. Frequency quadrature amplitude modulation (FQAM) has been shown to reduce the ICI at the cell edge therefore achieve a higher transmission rate for cell edge users. This paper investigates the detection of FQAM symbols and  noise plus  ICI in a multi-cell FQAM communication network. Turbo-coded bit error rate (BER) and frame error rate (FER) of multi-cell FQAM are studied. Also, the cumulative distribution function (CDF) of signal to noise plus interference (SINR) of multi-cell FQAM is computed using stochastic geometry. It is demonstrated via simulations FQAM outperforms quadrature amplitude modulation (QAM) in BER and FER when ICI is significant. Furthermore, FQAM can achieve better SINR than QAM.

\end{abstract}

\IEEEpeerreviewmaketitle


\section{Introduction}
One of the primary contributors to global mobile traffic growth is the increasing number of wireless devices that are accessing mobile networks. Each year, several million new devices with different form factors and increased capacities are being introduced. Over half a billion (526 million) mobile devices and connections were added in 2013 and the overall mobile data traffic is expected to grow to 15.9 exabytes per month by 2018, nearly an 11-fold increase over 2013~\cite{Cis14}. In addition to the large number of devices that need to access the network, emerging new services such as Ultra-High-Definition (UHD) multimedia streaming demand significantly increased cell capacity and end-user data rate \cite{whitepaper}. Such unprecedented growth in the number of connected devices and mobile data places new requirements \cite{Andrews14} for the fifth generation (5G) wireless access systems that are set to be commercially available around 2020.

In order to provide ubiquitous and high data rate connectivity, advanced small cells are envisaged for 5G \cite{whitepaper}. However, deployment of small cells with a higher density or smaller cell size in 5G causes a dilemma. On the one hand, the smaller the cells, the smaller the path loss, and therefore higher data rate is expected. On the other hand, such an advantage of increased data rate diminishes as having smaller cells introduces more severe inter-cell interference (ICI), which becomes one of the critical problems to solve in 5G. 

Frequency quadrature amplitude modulation (FQAM), considered as a combination of frequency shift keying (FSK) and quadrature amplitude modulation (QAM), can significantly improve transmission rates for cell-edge users \cite{Hong14, Hochwald03}. The mechanism of FQAM is that only one frequency component is actiave during each transmission period, over which a QAM symbol is transmitted. Information is conveyed by both the QAM symbol and the active frequency component index. The advantage of FQAM at cell edge comes from the fact that the statistics of aggregated ICI, created by transmitting FQAM symbols at the interfering BSs, is non-Gaussian, especially at the cell edge. As has been proved that the worst-case additive noise in wireless networks with respect to the channel capacity has a Gaussian distribution \cite{Seol09}, one can expect that the channel capacity can be increased by using FQAM. Variants of FQAM such as the generalized orthogonal frequency division multiplexing (OFDM) index modulation (IM) \cite{Fan15}, which activates multiple frequency components in each transmission period, and the generalized space and frequency IM \cite{Datta15}, which combines FQAM and spatial modulation (SM) \cite{Mesleh08}, have been reported in the literature.

Despite the significant advantages of FQAM and its potential of ICI reduction in 5G cellular networks, studies on FQAM has not drawn much attention in 5G. In this paper, based on \cite{Hong14}, we present and highlight the advantages of FQAM for 5G, comparing it with QAM. In particular, the detection of FQAM is studied, the noise plus ICI of FQAM under dense BS deployment is analyzed, and the cumulative distribution function (CDF) of signal to noise plus interference ratio (SINR) of multi-cell FQAM is derived using the stochastic geometry approach. The advantage of FQAM in terms of performance and SINR distribution is demonstrated and verified against simulation. 
 
The remainder of this paper is organized as follows. Section~\ref{System_model} gives the general description of the FQAM system. Section~\ref{sec_detection_algorithm} presents the detection, especially the computation of log-likelihood ratio (LLR) of Turbo-coded FQAM. In Section \ref{sec_ipn}, the noise pluse ICI of FQAM is analyzed and CDFs of SINR of FQAM are derived based on the statistic geometry approach, and are compared with those of QAM. Simulation results and comparisons are shown in Section \ref{sec_results_and_analysis} and conclusions are drawn in Section~\ref{conclusion_section}.
\section{System Model} \label{System_model}
\begin{figure} [!htb] 
\centering
\includegraphics[width=7cm]{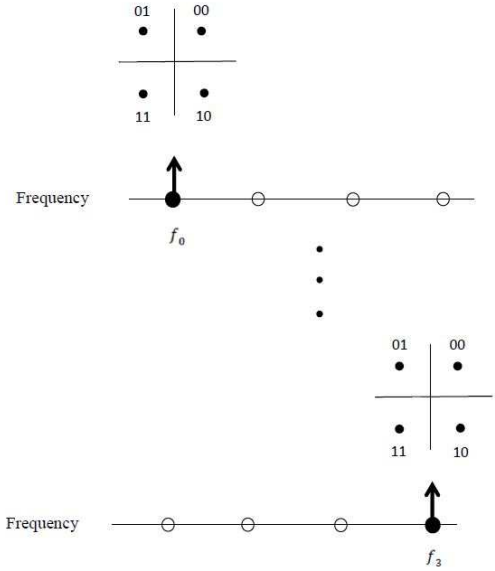}
\caption{Example of a $(4,4)$-FQAM signal constellation \cite{Hong14}.}
\label{fig1}
\end{figure}

We consider a homogeneous, synchronous, downlink cellular network with $N_B$ base stations (BSs). At each base station, a sequence of bits are interleaved, turbo-coded, and then modulated to FQAM symbols, which are used to transmit data over $N_s$ subcarriers. Assume $(M_\mathrm{F}, Q)$-FQAM symbols, which are formed by a combination of $M_\mathrm{F}$-ary FSK modulation and $Q$-ary QAM modulation, are used for transmission. It is known from \cite{Hong14} that a total of $(\log_2M_\mathrm{F}+\log_2Q)$ bits are mapped to one FQAM symbol, with the first $\log_2M_\mathrm{F}$ bits indicating the frequency index and the last $\log_2Q$ bits indicating the QAM index using Grey mapping.
 An example of a (4,4)-FQAM signal constellation is given in Fig.~\ref{fig1}.

\begin{figure}[t]
\centering
\includegraphics[width=4in]{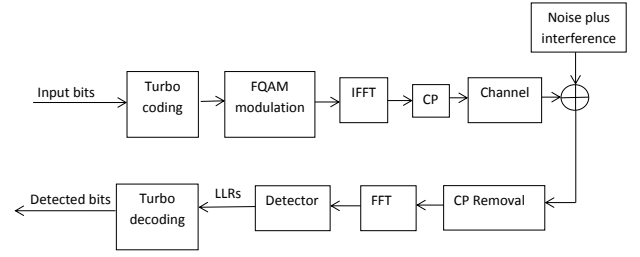}    
\caption{Diagram of the transceiver model of FQAM.}
\label{fig_FQAM_txrx}
\end{figure}
After FQAM modulation, identical to the QAM system, the length-$N_s$ FQAM symbols are processed with an inverse fast Fourier transform (IFFT), then a cyclic prefix (CP) is added at the beginning of the IFFT output, yielding one QAM symbol to be transmitted for each BS. These QAM symbols then go through respective fading channels from each base station to the user equipment (UE), where the channel between the $a$th ($a \in \{1,\cdots, N_B\}$) base station to the UE is given by a length-$L_a$ vector $\mathbf{h}^a = [h^a(0), \cdots, h^a(L_a-1)]^T$. At the receiver, the CP is removed, and a fast Fourier transform (FFT) is performed. It is known that the insertion and removal of CP together with the IFFT and FFT forms an equivalent one tap frequency domain channel on each subcarrier. The received signal is given by \cite{Hong14}
\begin{equation}\label{received}
\Omega_{k,l} = H_{k,l}^As_k^A\delta_{m_k^A, l} + \tilde{Z}_{k,l}
\end{equation}  
where $k$ ($k=0,\cdots, N_s-1$) is the frequency component index, $l$ ($l=0,\cdots, M_\mathrm{F}-1$) is the frequency index for the FQAM symbol at the $k$th frequency component, and $s_k^A$ represents the symbol transmitted on the $k$th frequency component at the desired BS, i.e., the $A$th base station, which takes a form of QAM symbol and $s_k^A\in\mathcal{S}$, where $\mathcal{S}$ denotes the set of the signals on a QAM constellation. In addition, $H_{k,l}^A$ is the frequency domain channel coefficient at the $k$th frequency component between the $A$th base station to the UE, given by taking the FFT of the time domain channel $\mathbf{h}^A$. Furthermore,  $m_k^A\in \{0,\cdots, M_\mathrm{F}-1\}$ is the frequency index of the FQAM symbol at the $k$th frequency component, and $\delta(\cdot)$ is the Dirac delta function. Finally, $\tilde{Z}_{k}$ is the corresponding noise plus ICI term, which includes the received signal from all other base stations $a = 0, \cdots, N_B-1$ and $a\neq A$. 

In order to detect the transmitted bits, a soft-decoding metric in the form of LLR is required to be computed as inputs to the turbo decoder. To obtain the soft-decoding metric, one could use the well-known maximum likelihood (ML) detector. However, such a detector assumes knowledge of the modulated symbols of the interfering BSs, which are practically unavailable at the receiver \cite{Hong14}. As a result, a  complex generalized Gaussian distribution (CGG) receiver based assuming CGG distribution of the noise plus ICI term is proposed in \cite{Hong14}, which we will detail in Section \ref{sec_detection_algorithm}. 

A block diagram of the transceiver of FQAM is detailed in Fig. \ref{fig_FQAM_txrx}. The transceiver structure of QAM is exactly the same as FQAM, except that in QAM, all frequency components are active. 
\section{Detection of FQAM} \label{sec_detection_algorithm}
The use of CGG detector in FQAM has been presented in literature \cite{Hong14}. In this section, we detail the process of a FQAM CGG detector for completeness.

It is known from \cite{Hong14} that assuming knowledge of the modulated symbols of the interfering BSs, one can use the conventional ML detector, considering the distribution of noise plus ICI as Gaussian. Such an assumption is however highly impractical. A sub-optimal detector was therefore proposed in \cite{Hong14}, assuming the CGG distribution of the noise plus ICI term. Such a suboptimal detector, namely a CGG detector, requires estimation of the shape and scale parameters, denoted as $\alpha$ and $\beta$ respectively, of the distribution of the noise plus ICI term.
\subsection{LLR computation for a CGG Detector}
LLR of a bit $b_k^\upsilon$ ($\upsilon = 0, \cdots, \log_2M_\mathrm{F}+\log_2Q-1$) of a CGG detector is given by \cite{Hong14}
\begin{align}\label{llr}
\mbox{LLR}_k = \mbox{ln}\frac{\sum_{(\tilde{m},\tilde{q})\in\tilde{B}_0^\upsilon}f_\mathbf{U}\left(\mathbf\Lambda_k^{(\tilde{m}, \tilde{q})}|\alpha, \beta\right)}{\sum_{(\tilde{m},\tilde{q})\in\tilde{B}_1^\upsilon}f_\mathbf{U}\left(\mathbf\Lambda_k^{(\tilde{m}, \tilde{q})}|\alpha, \beta\right)}
\end{align}
where $\tilde{B}_i^\upsilon$ denotes the set of all possible $(\tilde{m}\in\{0, \cdots, M_\mathrm{F}-1\}, \tilde{q}\in \{0, \cdots, Q-1\})$,  whose $v$th bit equals $i\in \{0,1\}$, and $\mathbf\Lambda_k^{\tilde{m}, \tilde{q}}$ is a length-$M_\mathrm{F}$ vector, with its $l$th ($l=0,\cdots, M_\mathrm{F}-1$) entry given by
\begin{equation}\label{lambdak}
\Lambda_k^{\tilde{m}, \tilde{q}}(l) = \Omega_{k,l}-H_{k,l}^As_{\tilde{q}}\delta_{\tilde{m},l}.
\end{equation} 
In addition, $f_\mathbf{U}\left(\cdot\right)$ is the joint probability density function (pdf) of $\mathbf{U}=[U_0, U_1, \cdots, U_{M_\mathrm{F}-1}]$, where $U_l = \tilde{Z}_{k,l}$ ($l=0,\cdots, M_\mathrm{F}-1$) is the independently and identically distributed (i.i.d.) random variables of the noise plus ICI term on the $k$th frequency component. The PDF is approximated as CGG distribution, given by\cite{Hong14} 
\begin{equation}\label{fu}
f_\mathbf{U}(\mathbf{u}|\alpha, \beta) = \left(\frac{\alpha}{2\pi\beta^2\Gamma(2/\alpha)}\right)^{M_\mathrm{F}}\prod_{l=0}^{M_\mathrm{F}-1}\exp\left(-\left(\frac{|u_l|}{\beta}\right)^\alpha\right)
\end{equation}
where $\alpha$, $\beta$ are the shape and scale parameters of the distribution. The estimation of $\alpha$ and $\beta$ is detailed in \cite[(21) and (22)]{Seol09}, which we give below for completeness. The $\alpha$ and $\beta$ are estimated as
\begin{equation}\label{alpha}
\hat{\alpha} = \frac{\eta}{\ln\left(\frac{(\sum|\hat{Z}_{k,l}|)^2}{N_s\sum|\hat{Z}_{k,l}|^2}-\xi\right)}+\ln\left(3/2\sqrt{2}\right)
\end{equation}
and
\begin{equation}\label{beta}
\hat{\beta} = \frac{\Gamma(2/\hat{\alpha})}{N_s\Gamma(3/\hat{\alpha})}\sum|\hat{Z}_{k,l}|
\end{equation}
where $\eta$ and $\xi$ are constants defined in \cite{Seol09}, 
the summation of $\hat{Z}_{k,l}$ is taken on all $k\in\{0,\cdots, N_s-1\}$ and $l\in\{0, \cdots, M_\mathrm{F}-1\}$, and $\hat{Z}_{k,l}$ is the estimated noise plus ICI term, given by \cite{Hong14}
\begin{equation}\label{estimation}
\hat{Z}_{k,l} = \Omega_{k,l} - H^A_{k,l}\hat{s}^A_k\delta_{\hat{m}_k^A, l}
\end{equation}
and 
\begin{equation}\label{minimization}
(\hat{m}_k^A, \hat{s}^A_k) = \arg\min_{{m_k^A\in \{0,\cdots, M_\mathrm{F}-1\}}\atop {s_k\in \{s_0, \cdots, s_{Q-1}\}}}\sum_{l=0}^{M_\mathrm{F}-1}\left|\Omega_{k,l}-H^A_{k,l}s^A_k\delta_{m_k^A,l}\right|^2.
\end{equation}

The algorithm of LLR computation for the CGG detector is given in Table \ref{LLRCGG}.  After obtaining the estimated $\alpha$ and $\beta$, substituting (\ref{lambdak}) and (\ref{fu}) to (\ref{llr}) yields the soft metric that is required for the subsequent turbo decoder, and the transmitted bits are detected followed by a deinterleaver. 

\begin{table}
\renewcommand{\arraystretch} {1.3}
\caption{Computing LLR of the CGG detector}
\label{LLRCGG}
\centering
\begin{tabular}{|l|}
\hline 
1: Solve the optimization problem given in (\ref{minimization})\\
2: Generate estimation of the noise plus ICI term according to (\ref{estimation}) \\
3: Estimate $\alpha$ and $\beta$ using (\ref{alpha}) and (\ref{beta}) \\
4: Obtain the pdf of noise plus ICI according to (\ref{fu}) \\
5: Compute LLR using (\ref{llr}) \\
\hline
\end{tabular}
\end{table}
We can further simplify the computation of LLR in step 5, by applying the maximum log approximation of LLR, given by
\begin{align}
\mbox{LLR}_k&=a\sum_{(\tilde{m},\tilde{q})\in \tilde{\mathcal{B}}_0^\upsilon}\prod_{l=0}^{M_\mathrm{F}-1}\exp\left(-\frac{|\Lambda_k^{\tilde{m}, \tilde{q}}(l)|}{\beta}\right)^\alpha\nonumber\\
&= \ln\frac{\sum_{(\tilde{m},\tilde{q})\in \tilde{\mathcal{B}}_0^\upsilon}\exp\left(-\sum_{l=0}^{M_\mathrm{F}-1}\frac{|\Lambda_k^{\tilde{m}, \tilde{q}}(l)|}{\beta}\right)^\alpha}{\sum_{(\tilde{m},\tilde{q})\in \tilde{\mathcal{B}}_1^\upsilon}\exp\left(-\sum_{l=0}^{M_\mathrm{F}-1}\frac{|\Lambda_k^{\tilde{m}, \tilde{q}}(l)|}{\beta}\right)^\alpha}\nonumber \\
&\approx-\frac{1}{\beta^\alpha}\left(\min\limits_{{(\tilde{m},\tilde{q})\in \tilde{\mathcal{B}}_0^\upsilon}}\left\lbrace \sum_{l=0}^{M_\mathrm{F}-1}\bigg|\Lambda_k^{\tilde{m}, \tilde{q}}(l) \bigg|^\alpha \right\rbrace \right. \nonumber\\
&\qquad\left. - \min\limits_{{(\tilde{m},\tilde{q})\in \tilde{\mathcal{B}}_1^\upsilon}}\left\lbrace \sum_{l=0}^{M_\mathrm{F}-1}\bigg|\Lambda_k^{\tilde{m}, \tilde{q}}(l) \bigg|^\alpha \right\rbrace\right)
\end{align}
where the last approximation comes from the well-known log-max approximation \cite{Hochwald03}. It is seen from the equation that when $\alpha=2$, Gaussian distribution is used to model noise plus ICI and the LLR is the same as that used in conventional ML detector for QAM. 


\section{Noise plus ICI analysis}\label{sec_ipn}
\subsection{FQAM vs QAM}
\begin{figure}[t]
\centering
\includegraphics[width=3.5in]{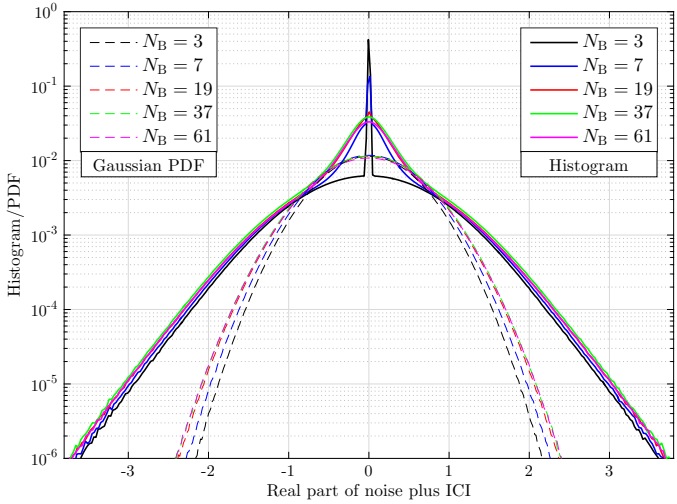}    
\caption{Histograms of the normalized real values of noise plus ICI samples in dense small cells network at the cell-edge region.}
\label{fig_ICI_Noise_Hist}
\end{figure}

The superiority of FQAM comparing to QAM is coming from the non-Gaussian distribution of the noise plus ICI. It has been shown that the noise plus ICI deviates from the Gaussian distribution in the macro cells environment \cite{Hong14}. However, it is expected that high dense small cells will dominate 5G systems. Thus, here we analyse the noise plus ICI for FQAM in high density small cells scenario. Fig. \ref{fig_ICI_Noise_Hist} shows the histogram for the real values of noise plus ICI at the cell-edge for different number of cells. Inter-site distance of 50m with 1 Watt transmission power are assumed. Apart from the three BSs case, the total number of BSs is based on the number of interference rings. The figure shows that in all cases the noise plus ICI distribution has much heavier tail compared to the Gaussian. The peak at the centre of the distribution is more prominent for small number of BSs (three and seven). The noise plus ICI distributions for higher number of base stations are almost identical. This indicate that even in a highly dense deployment, the noise plus ICI distribution for FQAM deviates from the Gaussian distribution. Hence, it is expected that the FQAM maintains its advantage comparing to QAM in dense deployment scenario.

To gain more insights of the performance difference between FQAM and QAM, in this section, we derive and compare the CDF of SINR of QAM and FQAM. In particular, we resort to the stochastic geometry approach, where BSs are assumed to be randomly located following a Poisson point process (PPP) with density $\lambda$. Such an assumption has been widely considered in the literature as a valid model which yields sufficient close analysis compared to that of practical models \cite{Bai14}\cite{Bai15}. We derive the CDF of SINR for FQAM in the following.

\subsection{SINR CDF analysis using stochastic geometry}
As there is only one frequency component that is active in FQAM. Therefore, the effective BS density is $\lambda/N_\mathrm{F}$. Let $\tilde{\rho}$ denote the SINR, $d$ denote the distance between the UE and the serving BS, $\sigma^2_N$ denote the noise power on each frequency component, and $\alpha$ denote the pathloss exponent. In addition, denote the set of all interfering BSs as $\tilde{A} \in \mathcal{B}\setminus A$, where $\mathcal{B}$ is the set of all BSs. The normalized interference power $\mathcal{I}$ of ICI can be computed as 
\begin{align}
\mathcal{I}=\sum_{\tilde{A}}|H^{\tilde{A}}|^2|d^{\tilde{A}}|^{-\alpha}
\end{align}
where the summation over $\tilde{A}$ is performed with respect to all interfering BSs. The channel $H^A$ between the target user and the serving BS follows Rayleigh distribution. Hence, $|H^A|^2$ is an exponentially distributed random variable. Let $\sigma^2_N$ be the noise power per frequency component, which can be computed via the noise power density $N_0$ (in dBm/Hz) and the bandwidth of each frequency component $W_\mathrm{sc}$ (in Hz) by $\sigma^2_N=10^{N_0/10}W_\mathrm{sc}\cdot 10^3$. Let $P_T$ denote the transmit power of BSs. Following the approach in \cite{Baccelli06}, the CDF $G^{\mathrm{FQAM}}_{\tilde{\rho}}(\tilde{\rho})$ of SINR can then be computed as
\begin{align}
&G^{\mathrm{FQAM}}_\mathrm{\tilde{\rho}}(\tilde{\rho})=1-\Pr\left\lbrace \frac{P_T|H^A|^2d^{-\alpha}}{\sigma^2_N+\mathcal{I}}\geqslant \tilde{\rho} \right\rbrace\nonumber\\
&=1-\exp\left(-d^\alpha\tilde{\rho}\sigma^2_N \right)\mathrm{E}\left[\exp\left(-d^\alpha \tilde{\rho} \mathcal{I} \right)\right]\nonumber\\
&=1-\exp\left(-d^\alpha P_T^{-1}\tilde{\rho}\sigma^2_N \right)\exp\left(-\frac{\lambda}{N_{\mathrm{F}}}d^2 \tilde{\rho}^{\frac{2}{\alpha}}\frac{2\pi^2}{\alpha\sin(2\pi/\alpha)} \right).
\label{equ_G_rho_FQAM}
\end{align}
Besides, for QAM, since all frequency components are active during transmission, the total noise power should be the summation over all frequency components. As a result, the CDF $G^{\mathrm{QAM}}_{\tilde{\rho}}(\tilde{\rho})$ of SINR can then be computed as \cite{Baccelli06} 
\begin{align}
&G^{\mathrm{QAM}}_\mathrm{\tilde{\rho}}(\tilde{\rho})\nonumber\\
&=1-\exp\left(-d^\alpha\tilde{\rho}N_{\mathrm{F}}\sigma^2_N \right)\exp\left(-\lambda d^2\tilde{\rho}^{\frac{2}{\alpha}}\frac{2\pi^2}{\alpha\sin(2\pi/\alpha)} \right).
\label{equ_G_rho_OFDM}
\end{align}
By comparing (\ref{equ_G_rho_FQAM}) and (\ref{equ_G_rho_OFDM}), it can be seen that QAM has a larger noise power due to larger active bandwidth. Additionally, QAM has larger ICI power than FQAM because all frequency components are active.

\section{Results and Analysis}\label{sec_results_and_analysis}
We present the simulation and numerical results in this section. First the performance of FQAM in terms of bit error rate (BER) and frame error rate (FER) is simulated. Then the numerical results on the CDF of SINR of FQAM are presented, and both the simulation and numerical results are compared with QAM. In all simulations, a multi-cell OFDM network and zero mean unit variance i.i.d. complex Gaussian channel are assumed.

\begin{figure}[t]
\centering
\includegraphics[width=3.5in]{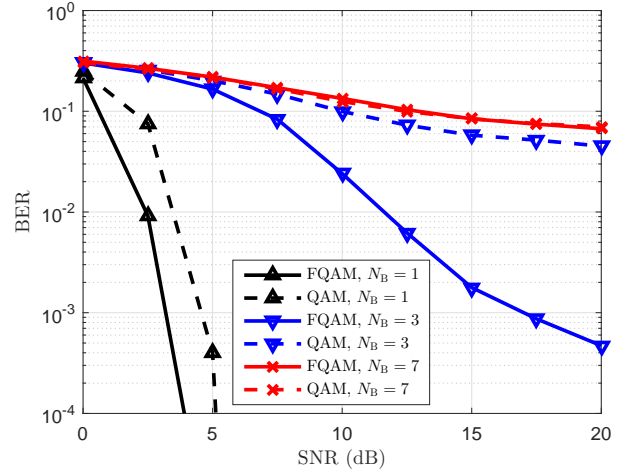}    
\caption{BER of FQAM and QAM with different numbers of BSs ($N_\mathrm{F}=4$, $1$ bit/frequency component).}
\label{fig_BER_FQAMvsOFDM}
\end{figure}

\begin{figure}[t]
\centering
\includegraphics[width=3.5in]{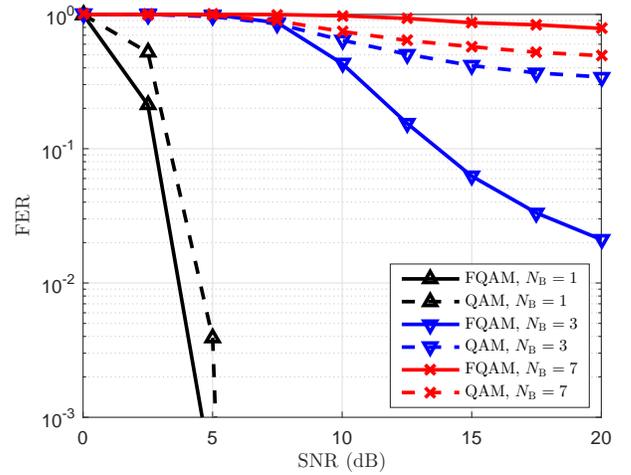}    
\caption{FER of FQAM and QAM with different numbers of BSs ($N_\mathrm{F}=4$, $1$ bit/frequency component).}
\label{fig_FER_FQAMvsOFDM}
\end{figure}

BER and FER comparisons between FQAM and QAM with respect to different numbers of BSs are depicted in Fig. \ref{fig_BER_FQAMvsOFDM} and Fig. \ref{fig_FER_FQAMvsOFDM}, respectively. In these simulations, $1/3$ code rate Turbo code is used. The location of the UE is assumed to be at the cell edge of the serving BS and in the center of three closest BSs for $N_\mathrm{B}=3$ and $N_\mathrm{B}=7$, which is essentially the worst case scenario of ICI for users in cellular networks. To have fair comparison, both FQAM and QAM have the same spectral efficiency, i.e., $1$ bit/frequency component. It can be observed that FQAM outperformed QAM in terms of BER and FER with single or three BSs. For $N_\mathrm{B}=1$, the gain of FQAM comes from the higher SNR per frequency component as FQAM allocates all power on the only one active frequency component while QAM allocates its power on all active frequency components. The gap between FQAM and QAM becomes more significant with three BSs because less interference is received in FQAM when only one frequency component is active. When the number of BSs reaches seven, neither FQAM nor QAM performs well due to the ICI.

\begin{figure}[t]
\centering
\includegraphics[width=3.5in]{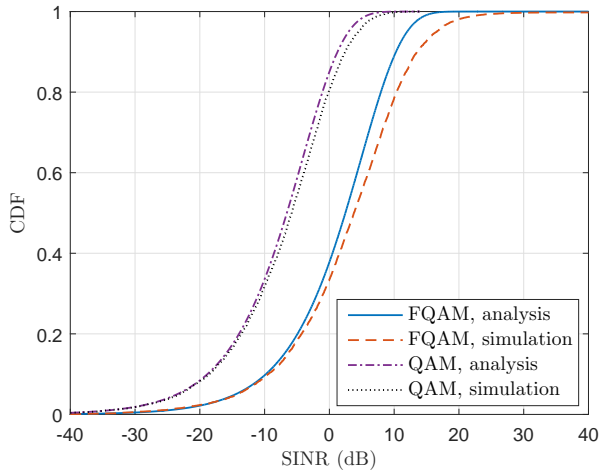}    
\caption{SINR CDFs of FQAM and QAM ($N_\mathrm{F}=4$, $\lambda=10^{-4}$, $\alpha=3$, $d=50$m, $N_0=-173$dBm/Hz, $W_\mathrm{sc}=15000$Hz, $P_T=20$W).}
\label{fig_SINR_CDF_FQAMvsOFDM}
\end{figure}

SINR CDFs of FQAM and QAM are compared in Fig.~\ref{fig_SINR_CDF_FQAMvsOFDM}. It can be observed that analysis results based on stochastic geometry fits the simulation well. Also, the SINR of QAM systems is smaller than that of FQAM, where a difference of around 10 dB is observed between two medians. This is because FQAM introduces randomness in the frequency domain to reduce ICI. 


\section{Conclusions} \label{conclusion_section}
This paper has presented the performance of FQAM in terms of BER and FER under interference scenarios, and compared with that of QAM. In addition, the CDF of SINR for FQAM is also analysed, numerically computed, and compared with that of QAM. The advantage of FQAM over QAM in terms of BER and FER at cell edge for both single and multiple BS scenarios has been demonstrated.  In particular, significant performance gain has been shown with a reasonably practical scenario where $N_B=3$ BSs is considered. Advantage of FQAM in terms of the distribution of SINR has also been shown, where a SINR difference of around 10 dB is observed at an outage of 10\%. All these advantages suggest that much more attention should be raised in considering FQAM as a promising technology in the 5G mobile networks.

\section*{Acknowledgment}
This work has been performed in the framework of the Horizon 2020 project FANTASTIC-5G (ICT-671660) receiving funds from the European Union. The authors would like to acknowledge the contributions of their colleagues in the project, although the views expressed in this contribution are those of the authors and do not necessarily represent the project. The authors would also like to thank Sungnam Hong from Samsung Electronics for his insightful suggestions.


\begin{thebibliography}{1}
\bibitem{Cis14} Cisco White Paper, ``Cisco Visual Networking Index: Global
Mobile Data Traffic Forecast Update," Feb. 2014. Available:
\\
http://www.cisco.com/c/en/us/solutions/service-provider/visual-networking-index-vni/white-paper-listing.html
\bibitem{whitepaper} Samsung White Paper, ``5G Vision", Feb. 2015. Available:
\\
http://www.samsung.com/global/business-images/insights/2015/Samsung-5G-Vision-0.pdf



\bibitem{Andrews14} J. G. Andrews, S. Buzzi, W. Choi, S. V. Hanly, A. Lozano, A. C. K. Soong, and J. C. Zhang, ``What will 5G be?," \emph{IEEE J. Sel. Areas Commun.}, vol. 32, no.~6, pp. 1065--1082, June 2014.

\bibitem{Hong14} S. Hong, M. Sagong, C. Lim, S. Cho, K. Cheun, and K. Yang, ``Frequency and quadrature-amplitude modulation for downlink cellular OFDMA networks," \emph{IEEE J. Sel. Areas Commun.}, vol. 32, no.~6, pp. 1256--1267, June 2014.

\bibitem{Hochwald03} B. M. Hochwald and S. Ten Brink, ``Achieving near-capacity on a multiple-antenna channel," \emph{IEEE Trans. Commun.}, vol. 51, no. 3, pp. 389--399, Mar. 2003.

\bibitem{Seol09} C. Seol and K. Cheun, ``A statistical inter-cell interference model for downlink cellular OFDMA networks under log-normal shadowing and multipath Rayleigh fading," \emph{IEEE Trans. Commun.}, vol. 57, no. 10, pp. 3069--3077, Oct. 2009.



\bibitem{Fan15} R. Fan, Y. J. Yu, and Y. L. Guan, ``Generalization of orthogonal frequency division multiplexing with index modulation," \emph{IEEE Trans. Wireless Commun.}, vol. 14, no. 10, pp. 5350--5359, Oct. 2015.


\bibitem{Datta15} T. Datta, H. S. Eshwaraiah, and A. Chockalingam, ``Generalized space and frequency index modulation," \emph{IEEE Trans. Veh. Technol.}, accepted.


\bibitem{Mesleh08} R. Y. Mesleh, H. Haas, S. Sinanovic, W. A. Chang, and Y. Sangboh, ``Spatial modulation," \emph{IEEE Trans. Veh. Technol.}, vol. 57, no. 4, pp. 2228--2241, July 2008.


\bibitem{Baccelli06} F. Baccelli, B. Blaszczyszyn, and P. Muhlethaler, ``An Aloha protocol for multihop mobile wireless networks," \emph{IEEE Trans. Inf. Theory.}, vol. 52, no. 2, pp. 421--436, Feb. 2006.

\bibitem{Baccelli_v1} F. Baccelli and B. Blaszczyszyn, \emph{Stochastic Geometry and Wireless Networks Volume I: Theory.}, Foundations and Trends\textregistered in Networking: vol. 3: no. 3--4, pp. 249--449, 2010.

\bibitem{Baccelli_v2} F. Baccelli and B. Blaszczyszyn, \emph{Stochastic Geometry and Wireless Networks Volume II: Applications.}, Foundations and Trends\textregistered in Networking: vol. 4: no. 1--2, pp. 1--312, 2010.

\bibitem{Bai14} T. Bai, A. Alkhateeb, R. W.  Heath Jr, ``Coverage and capacity of millimeter-wave cellular networks," \emph{IEEE Commun. Mag.}, vol. 52, no. 9, pp. 70--77, Sep. 2014.

\bibitem{Bai15} T. Bai and R. W. Heath Jr, ``Coverage and rate analysis for millimeter-wave cellular networks," \emph{IEEE Trans. Wireless Commun.}, vol. 14, no. 2, pp. 1100--1114, Oct. 2014.
%
%
%



\bibitem{ElementsIT} T. M. Cover and J. A. Thomas, \emph{Elements of Information Theory.}, 2nd ed., Wiley \& Sons, New Jersey, 2006.



\end{thebibliography}
\end{document}